\documentstyle[preprint,prc,eqsecnum,aps]{revtex}
\begin{document}

\def\lsim{\thinspace{\hbox to 8pt{\raise
 -5pt\hbox{$\sim$}\hss{$<$}}}\thinspace}
\def\rsim{\thinspace{\hbox to 8pt{\raise
 -5pt\hbox{$\sim$}\hss{$>$}}}\thinspace}

\title{The Role of Final State Interactions in Quasielastic \\
 $^{56}$Fe$(e,e')$ Reactions at large $|\vec q|$ }

\author{ C.R.~Chinn}
\address{ Department of Physics and Astronomy \\
Vanderbilt University,
Nashville, TN  37235}

\maketitle

\vspace{1in}

\begin{abstract}
A relativistic finite nucleus calculation using a
Dirac optical potential is used to investigate the
importance of final state interactions [FSI]
at large momentum transfers in inclusive quasielastic
electronuclear reactions.  The optical potential is derived from
first-order multiple scattering theory and then is used to calculate
the FSI in a nonspectral Green's function
doorway approach.  At intermediate momentum transfers
excellent predictions of the quasielastic $^{56}$Fe$(e,e')$
experimental data for the longitudinal response function
are obtained.  In comparisons with recent
measurements at $|{\vec q|}=1.14$~GeV/c the theoretical calculations
of $R_L$ give good agreement for the quasielastic peak shape and
amplitude, but place the position of the peak at an energy
transfer of about $40$~MeV higher than the data.
\end{abstract}

\vspace{0.5in}

\pagebreak


\narrowtext


\section{Introduction}

There has been a recent interest in inclusive quasielastic
electronuclear reactions at large momentum transfers
($|\vec q| \rsim 1$~GeV/c), especially with the upcoming programs
planned at \mbox{CEBAF}.  Along with recent measurements \cite{Chen}
there has been some theoretical work investigating the physical role
of relativity and final state interactions [FSI] in such reactions
\cite{Frank1,Frank2}.  From these studies it has been noted that a
full consistent finite nucleus calculation would be helpful in
discerning the various physical contributions.  This paper attempts to
address that need.

In the simple relativistic Fermi gas
calculation of Ref.~\cite{Frank1} the
implications were that the role of FSI at $|\vec q| \rsim 1$~GeV/c
appeared to be greatly reduced, especially since the use of a real
energy-independent potential to model the FSI caused the predicted
position of the quasielastic peak to move significantly away from
the data.  The use of a Fermi gas model
may be misleading however, since the recoil effects will be
misrepresented which may affect the calculated position of
the quasielastic peak.  A finite nucleus model should be more
appropriate.  In Ref.~\cite{Frank2} an energy-dependent
real potential was introduced to give the FSI, and the conclusion
from this study is that the FSI remain important at large
$|\vec q|$, and that the energy dependence is required to enable
one to predict the peak position correctly.  In that work the imaginary
part of the optical potential is neglected.  From optical model
studies of elastic nucleon-nucleus scattering and from multiple
scattering theory there is known to be a strong energy dependence
in the optical potentials.   To delete the imaginary part of an
energy-dependent potential is to break the unitarity constraint
and thus incorrectly represent the reactive content of the
optical potential.  For a more physical representation, the
full complex energy-dependent optical potential should be
included in a consistent manner.  The importance of such
considerations was discussed in detail in Ref.~\cite{chinn1}.

In this paper two complex energy-dependent Dirac optical potentials
derived from multiple scattering are used in a relativistic
finite-nucleus calculation to calculate the separated response
functions for inclusive quasielastic $(e,e')$ scattering from
$^{56}$Fe.  In these calculations Dirac dynamical
effects resulting from couplings to negative energy states, which were
shown in Ref.~\cite{chinn1} to be important, are
included.  In Section~II a theoretical discussion of the model and
the calculation is presented.  The results and the comparison with
the experimental measurements are presented in Section~III followed
by a conclusion.

\vspace{1cm}

\section{Theoretical Discussion}

Although the main mechanism in quasielastic reactions is assumed
to be the knockout of a single nucleon, in inclusive reactions all
possible final states are included in the experimental measurements.
The optical potential implies the existence of other final states
besides the knockout channel within the imaginary part.
For this reason to include all of the possible final states
implied by an optical potential the nonspectral Green's function
doorway approach \cite{hlm}, which is discussed more fully in
Ref.~\cite{chinn1}, is used.
The longitudinal and transverse response functions within the
one-photon-exchange approximation are given by:
\begin{eqnarray}
R_L({\vec q},\omega) & = & W^{00}({\vec q},\omega) \nonumber \\
R_T({\vec q},\omega) & = & W^{11}({\vec q},\omega) +
                           W^{22}({\vec q},\omega) ,
\end{eqnarray}
where
\begin{equation}
W^{\mu \nu} = {\overline {\sum_i}}\int{\hskip -16pt}\sum_f
              \langle i| {\widehat J}^{\mu}(q)^\dagger | f\rangle
              \delta(E_i + \omega - E_f )
    \langle f| {\widehat J}^{\nu}(q) | i \rangle .  \label{eq:2.1}
\end{equation}
Here $| i \rangle$ represents the initial nuclear many-body state,
while the sum over $| f \rangle$ corresponds to all final states of
the full hadronic many-body assembly.  ${\widehat J}^{\mu}(q)$ is
the electromagnetic nuclear current operator and the
$\overline {\sum_i}$ denotes an average over the initial states.

One would like to perform an explicit sum over the complete set of
complex inelastic reaction channels in the final state, but in
practice such a many-body calculation is prohibitively difficult.
Therefore a nonspectral approach is used where the sum over all
of the final states within a particular space is implicitly
performed, by considering the full A-body Green's function.

Suppressing the discrete state contribution, eq.~(\ref{eq:2.1}) can
be rewritten in terms of the forward virtual Compton amplitude:
\begin{equation}
W^{\mu\nu}({\vec q},\omega) = - \frac{1}{\pi} Im
                         T^{\mu\nu} ({\vec q},\omega) ,
\end{equation}
where
\begin{equation}
T^{\mu\nu} ({\vec q},\omega) = {\overline {\sum_i}}
    \langle i | {\widehat J}^{\mu}(q)^{\dagger}
{\widehat G}(\omega+E_i) {\widehat J}^{\nu}(q)
 | i \rangle .  \label{eq:2.3}
\end{equation}
Here ${\widehat G}$ is the full many-body propagator for the
$A-$nucleon system.  If ${\widehat J}^{\mu}$ is assumed to be
a one-body operator, it can be shown \cite{chinn1} that within
this one-body space ${\widehat G}$ reduces to the optical model
Green's function.  By using the optical model Green's function
in a nonspectral form, then a proper and consistent unitary
description of the reactive content of this inclusive reaction
is maintained \cite{chinn1}.  If one assumes that each knockout
channel is represented by the same optical model potential, then
the following substitution can be made:
\begin{equation}
{\widehat G} \approx {\widehat G}_{opt} = {\widehat G}_0 +
   {\widehat G}_0 ~ {\widehat U}_{opt} ~ {\widehat G}_{opt}.
\end{equation}
${\widehat G}_{opt}$ corresponds to the use of an optical model
potential to represent the final state interactions between the
ejected nucleon and the residual nucleus.  ${\widehat G}_0$ is the
free propagator for a nucleon within the nuclear medium.

To reduce the calculation to the plane wave approximation
[PWA], ${\widehat U}_{opt}$ is set to zero or equivalently:
\begin{equation}
{\widehat G} \approx {\widehat G}_0 ,
\end{equation}
which leads to eq.~(\ref{eq:2.1}), where only final plane wave
states for the ejected nucleon are considered.

The calculations are performed in a fully-off-shell momentum space
representation:
\begin{equation}
T^{\mu\nu}({\vec q},\omega) = {\overline {\sum_i}} \int
\frac{d^3{\vec p}\,d^3{\vec p'}}{2\pi^3} \langle{\bar i}|
 {\vec p}-{\vec q}\rangle\, J^\mu(-q) \,
  G_{opt}({\vec p},{\vec p'};E) \, J^\nu(p') \,
 \langle{\vec p'}-{\vec q}| i\rangle ~. \label{eq:2.4}
\end{equation}
Here $G_{opt}({\vec p},{\vec p'};E)$ is calculated from the fully
off-shell relativistic optical potential as derived from multiple
scattering theory.  The optical model Green's
function is calculated as the solution of the Lippmann-Schwinger
equation in momentum space to give the fully off-shell
nucleon-nucleus $T$-matrix.  The equations are solved
in partial wave form so as to include the spin-orbit contributions
in a convenient way.
\begin{eqnarray}
G_{opt}({\vec p},{\vec p'};E) & = & G_0(p;E)\,\delta^{(3)}
 ({\vec p}-{\vec p'}) + G_0(p;E)\,
 V_{opt}({\vec p},{\vec p'})\, G_{opt}({\vec p},{\vec p'};E) \\
 & = & G_0(p;E)\,\delta^{(3)}({\vec p}-{\vec p'}) + G_0(p;E)\,
   T_{opt}({\vec p},{\vec p'};E)\, G_0(p';E) \nonumber \\
T_{opt}({\vec p},{\vec p'};E) & = & V_{opt}({\vec p},{\vec p'}) +
 \int d^3{\vec p''}\, V_{opt}({\vec p},{\vec p''}) \,G_0(p'';E) \,
  T_{opt}({\vec p''},{\vec p'};E)
\end{eqnarray}
In the above equation the free Dirac propagator can be separated
into a positive-energy projecting part and a negative-energy
projecting part, so that contributions to eq.~(\ref{eq:2.4})
involve coupling to the negative-energy Dirac sea.  It has
been shown that such effects can play a major role in the
calculated response functions \cite{chinn1,chinn2,chinn3,chinn4}.

In the PWA and FSI calculations a single particle description is
used.  Bound state wave functions are taken from a Dirac-Hartree
calculation \cite{hs} and are represented in Dirac 4-spinor form.
Since the Dirac-Hartree calculation performed in Ref.~\cite{hs}
assumes spherical symmetry and $^{56}$Fe is not a doubly magic
closed shell nuclei, approximations are used to represent the
valence nucleons.  In this case the valence shell is represented as a
closed shell, but with fractional occupation numbers to give 26
protons and 30 neutrons.  The current operators are treated in
relativistic form and no nonrelativistic reduction is performed.

The analysis presented in this paper will
be performed for both a.) the relativistic plane wave
approximation and b.) with the Green's function doorway
approach to include the FSI.  The form of the electromagnetic
nucleon current inside of the nucleus is unknown, hence two
different functional forms of the free electromagnetic current
operator are used:
\begin{eqnarray}
{\widehat J}_{cc2}^{\mu} & = & F_1(q^2) \gamma^{\mu} + \imath
\frac{F_2(q^2)}{2 m} \sigma^{\mu\nu} q_{\nu} , \nonumber \\
 {\rm and} & ~ & \label{eq:2.5}\\
{\widehat J}_{cc1}^{\mu} & = & G_m(q^2) \gamma^{\mu} -
                           \frac{F_2(q^2)}{2 m} {\bar {\cal K}}^{\mu}
                           \nonumber \\
                     & = & F_1(q^2) \gamma^{\mu} + \imath
\frac{F_2(q^2)}{2 m} \sigma^{\mu\nu} {\bar q}_{\nu} , \nonumber
\end{eqnarray}
where $G_m = F_1 + F_2 $ is the familiar Sachs magnetic form factor
and ${\cal K} \equiv k + k'$.  The bars over
${\bar {\cal K}}$ and ${\bar q}$ indicate that $k$ and $k'$ are
fixed to the onshell values, {\it e.g.}
$\bar k^o \equiv \pm \sqrt{{\vec k}^2 + m^2}$, where the sign is
dependent upon the $(\pm)$ energy character of the Dirac spinor.
The definitions of ${\widehat J}_{cc2}$ and ${\widehat J}_{cc1}$
correspond to the $cc2$ and
$cc1$ operators defined in Ref.~\cite{DeF}.  These two operators
are identical on-shell and hence give the same free nucleon
electromagnetic representations.  In the off-shell case
where one scatters from bound nucleons or when one includes FSI,
these operators give differing results.  The most general form
of the current operator contains 12 independent terms, in which
only two independent terms survive in the on-shell limit.  To be
able to construct the complete operator with the accompanying form
factors would require a reliable off-shell
nucleon structure model, for example a QCD based model.  Hence the
calculated differences between the $cc1$ and $cc2$ operators can
only be understood in terms of the underlying nucleon structure.
For a detailed analysis of
the effects and uncertainties represented by these two operators,
please see Ref.~\cite{chinn3,rNKa,rTT}.

Current conservation is imposed by means of the standard replacement of
${\widehat q}\cdot {\vec J}$ by $q_o J^o/|{\vec q}|$.  In general
the current is not conserved by these two operators, since there
typically is not a consistent Hamiltonian treatment of the initial
state, the final state and the electromagnetic current interactions.
The form factors used in this paper are taken from Ref.~\cite{gk}.

Two relativistic complex optical potentials derived from multiple
scattering theory are used to represent the FSI.  In this case
nucleon-nucleon [$NN$] t-matrices are folded with local densities in
the optimum factorization approximation \cite{pttw} to give the
optical potential.  The negative-energy part of the optical
potential is constructed using an approximate approach \cite{hptt}.
For one optical potential the $NN$ t-matrices are calculated from
the fully-off-shell full Bonn potential \cite{bonn}, which include the
effects of relativistic kinematics, retarded meson propagators as
given by time-ordered perturbation theory, and crossed and
iterative meson exchanges with $NN$, $N\Delta$ and $\Delta\Delta$
intermediate states.  For ejectile energies greater than 300~MeV,
an extension of the Bonn meson exchange interaction above pion
production threshold is used \cite{bonn2}.  The second optical
potential uses the $NN$ interaction of Ref.~\cite{fl85}.
The proton densities are taken from electron scattering
measurements \cite{electron}, while the neutron densities are
those calculated from the Hartree-Fock-Bogolyubov calculation of
Ref.~\cite{HFB}.  In calculations of elastic nucleon-nucleus scattering,
the use of the Bonn potential tends to give a better representation
of the data than the Franey-Love amplitudes, probably due to the
superior off-shell behavior of the Bonn potential.

In this paper the nonrelativistic calculations of FSI are constructed
from the relativistic calculations with the exception that all of
the negative energy contributions which result from the Dirac
dynamics are neglected.
This includes those negative energy contributions that
arise from the construction of ${\widehat U}_{opt}$ and in the
calculation of ${\widehat G}_{opt}$.  This is the manner in which
the nonrelativistic calculation is calculated here,
where relativistic kinematics are maintained.

\section{Results and Comparisons with the Data}

To gauge the accuracy of the theoretical model, comparisons with
quasielastic $^{56}$Fe$(e,e')$ data are made at $|{\vec q}|=$~410
and 550~MeV/c.  These comparisons are made both with
nonrelativistic FSI and with relativistic FSI including Dirac
dynamical degrees of freedom.  The PWA results with no FSI
$(V_{opt}=0)$ and the nonrelativistic FSI calculations are shown
in Figs.~1 and 2 for $|{\vec q}|=$~410 and 550~MeV/c, respectively.
The PWA calculation places the peak position for $R_L$ in the
upper panels at an energy transfer of about $15-25$~MeV larger
than the experimental data.  Note that these data do not include
any Coulomb distortion corrections, which may shift the
experimental result.  After including the nonrelativistic FSI, one
can see the peak position is much better represented with the
error in $R_L$ being negligible.

The use of the Bonn potential in the solid curve gives a slightly
different result from the long-dashed curve, which uses the
Franey-Love amplitudes.  In this case the Bonn potential
gives a slightly better representation of the data.

In the middle and lower panels of Figs.~1 and 2 the transverse
response is calculated.  In this case there are two possible
predictions for the same data due to the ambiguity about the
${\widehat J}_{cc1}$ and ${\widehat J}_{cc2}$ current operators in
eq.~(\ref{eq:2.5}).  These two operators give formally identical
longitudinal results, but differ in the transverse channel.  For
the nonrelativistic case this difference is very small.  Since the
$\Delta$~resonance is not included in this calculation the
comparisons with the data can only be made qualitatively.  From
the tail of the $\Delta$~resonance in the curves, it appears that
the nonrelativistic FSI predictions will underestimate the data.

The relativistic FSI results for $|{\vec q}|=$~410 and 550~MeV/c are
shown in Figs.~3 and 4, respectively.  Here the predictions of
$R_L$ are very close, where the Bonn potential gives a
slightly larger $R_L$ than the Franey-Love amplitudes.  The peak
positions are accurate and the overall peak is well-represented.
For the transverse case the $cc2$ or Dirac current results shown
in the middle panels greatly underestimate the data, while the
$cc1$ current results are very close to the data, although the
inclusion of the $\Delta$-resonance degrees of freedom may easily
alter this agreement.

{}From the $R_L$ comparisons the relativistic dynamical effects are
very important in providing good theoretical predictions of the
data at these intermediate momentum transfers.  For greater
momentum transfers one would expect that relativistic effects to
be an important and necessary ingredient for any accurate
theoretical description of the data.

The quasielastic $^{56}$Fe$(e,e')$ results for a momentum transfer
of $|{\vec q}|=1.14$~GeV/c are shown in Figs.~5 and 6 using
nonrelativistic and relativistic FSI, respectively.  Here the PWA
result for $R_L$ places the peak positions at an energy transfer
which is about 55~MeV greater than the peak in the data.  With
nonrelativistic FSI, the peak position is closer to the data but
still about $30-45$~MeV too high.  With relativistic FSI in Fig.~6
the Dirac dynamical effects move the predicted peak further away
to be about $40-50$~MeV higher in $\omega$ than the data.  In this
case the shapes of the peak in $R_L$ are well-represented,
although a bit too wide, while the peak position is
not as accurately placed.  The relativistic FSI calculation gives
a smaller amplitude peak than the nonrelativistic FSI calculation
and is closer to the circled data, which take into account in an
approximate fashion the Coulomb distortions.
It is interesting to note that the relatively large errors are not
truly able to discern clearly between the relativistic and
nonrelativistic FSI calculations.  The Coulomb corrected data appear
to favor the need for relativistic FSI.
The predicted peak position in Fig.~6 is actually closer than
the most sophisticated results of Ref.~\cite{Frank2}, which places
the peak position at
an $\omega$ of about 85 MeV higher than the data.  It is clear
from Fig.~6 than FSI are very important effects at even this high
momentum transfer, although one would like to obtain a more
accurate prediction of the peak position.

In some sense in Fig.~6, $40$~MeV does not seem like a large
number, but on the scale of Figs.~3 and 4 this value becomes
significant.  The incorrect peak position seen in Fig.~6 cannot be
interpreted in terms of an average binding energy shift, since
this would also affect the intermediate energy range results by
the same amount, where the peak positions are accurately
reproduced.  The shift in the peak position must arise from a
dynamical effect.  There are a number of possible candidates for
such an effect, such as the restoration of current conservation, meson
exchange effects, ambiguities in the off-shell structure of the
electromagnetic current, FSI effects not included in an optical
model description, better treatment of the Coulomb distortions
or even the need for a fully causal or Lorentz invariant
description.  With the advent CEBAF, comparisons with other
more accurate experimental results, especially in the $(e,e'p)$
case, may prove to be very enlightenning.

\section{Conclusion}

The longitudinal and transverse response functions for the
inclusive quasielastic electronuclear scattering reaction from
$^{56}$Fe are calculated using a relativistic finite nucleus model
with FSI and Dirac dynamical degrees of freedom.  FSI
are included using the optical model Green's function doorway
formalism, which for
the case of a one-body electromagnetic current operator provides a
consistent description of the final states resulting from the reactive
content implied by the imaginary part of the optical potential.
The optical model Green's function is calculated in a
fully-off-shell momentum space calculation using optical
potentials derived from first-order multiple scattering theory
using two different nucleon-nucleon interactions.

It is found that
at intermediate momentum transfers of 410 and 550~MeV/c, the
relativistic FSI calculation gives a very good theoretical
description of the data, reproducing well the position, shape and
amplitude of the quasielastic peak.  At $|{\vec q}|=1.14$~GeV/c
comparisons with recent data find that the shape and amplitude of
the peak are well-produced, but that the peak is placed at an energy
transfer of about $40$~MeV higher than the experimental result.
The source of this discrepancy is not clear and may provide a
motivation for future investigations.

\vfill

\acknowledgments
Appreciation is extended to J.~P.~Chen and Z.~E.~Meziani for
sharing their data and to R.~M.~Thaler for helpful discussions.
This work was performed in
part under the auspices of the U.~S. Department of Energy under
contracts No. DE-AC05-84OR21400 with Martin Marietta Energy
Systems, Inc., and DE-FG05-87ER40376 with Vanderbilt University.
This research has been supported in part by the U.S. Department
of Energy, Office of Scientific Computing under the High
Performance Computing and Communications Program (HPCC) as a
Grand Challenge project titled ``the Quantum Structure of Matter.''


\pagebreak


\pagebreak

\begin{figure}
\caption{The inclusive quasielastic separated response functions
        are shown for scattering from $^{56}$Fe at
        $|{\vec q}|=410$~MeV/c.  The longitudinal response is shown
        in the upper panel.  The transverse response function
        calculated using the $cc2$ (Dirac) and the $cc1$
        electromagnetic nucleon current operators is shown in the
        middle and lower panels, respectively.  The short-dashed
        curves represent the PWA calculation.  The solid and
        long-dashed curves correspond to the nonrelativistic FSI
        calculation using optical potentials calculated with the
        full Bonn potential and Franey-Love amplitudes,
        respectively.  The data do not include any Coulomb
        distortion corrections and are from Ref.~\protect\cite{Meziani}. }
\end{figure}

\begin{figure}
\caption{The same as Fig. 1, except at $|{\vec q}|=550$~MeV/c. }
\end{figure}

\begin{figure}
\caption{The same as Fig. 1, except the FSI are calculated using
        the relativistic model including Dirac dynamical effects. }
\end{figure}

\begin{figure}
\caption{The same as Fig. 1, except at $|{\vec q}|=550$~MeV/c and
        the FSI are calculated using the relativistic model
        including Dirac dynamical effects. }
\end{figure}

\begin{figure}
\caption{The same as Fig. 1, except at $|{\vec q}|=1.14$~GeV/c and
        the data are from Ref.~\protect\cite{Chen}. The data
        represented by the circles accounts for Coulomb distortion
        effects by using an effective $q$, while the squares omit
        this correction.  }
\end{figure}

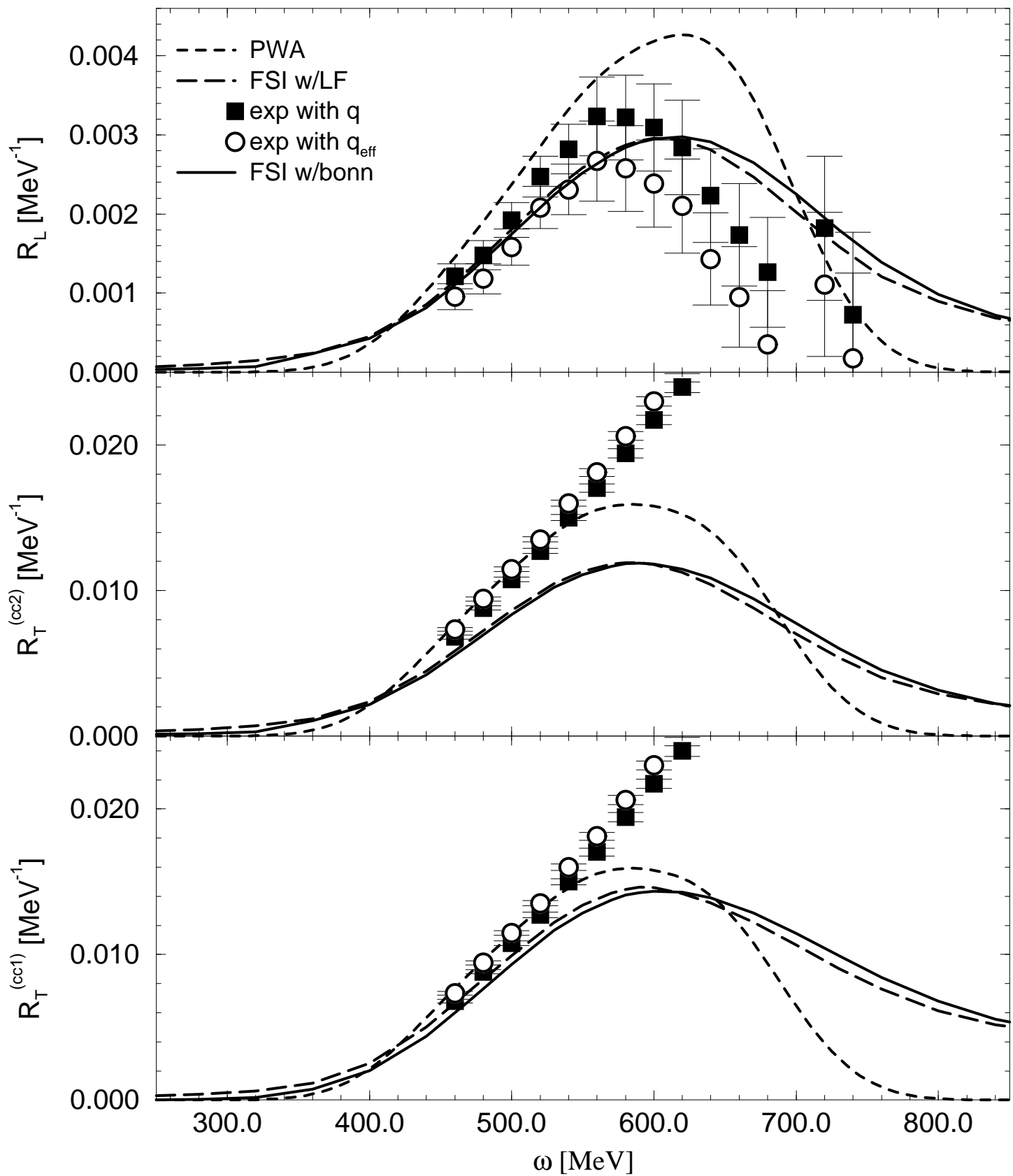
\begin{figure}
\caption{The same as Fig. 1, except at $|{\vec q}|=1.14$~GeV/c and
        the FSI are calculated using the relativistic model
        including Dirac dynamical effects and the data are from
        Ref.~\protect\cite{Chen}.  The data represented by the
        circles accounts for Coulomb distortion effects by using an
        effective $q$, while the squares omit this correction.  }
\end{figure}

\end{document}